# The spatial structure of cell signaling systems

**Ruth Nussinov**[1,2,*]

[1]*Basic Research Program, SAIC-Frederick, Inc.*
*Cancer and Inflammation Program,*
*National Cancer Institute,*
*Frederick, MD 21702*

[2]*Sackler Inst. of Molecular Medicine*
*Department of Human Genetics and Molecular Medicine*
*Sackler School of Medicine, Tel Aviv University, Tel Aviv 69978, Israel*

Ruth Nussinov, NussinoR@helix.nih.gov

**Abstract**

The spatial structure of the cell is highly organized at all levels: from small complexes and assemblies, to local nano- and micro-clusters, to global, micrometer scales across and between cells. We suggest that this multiscale spatial cell organization also organizes signaling and coordinates cellular behavior. We propose a new view of the spatial structure of cell signaling systems. This new view describes cell signaling in terms of dynamic allosteric interactions within and among distinct, spatially organized transient clusters. The clusters vary over time and space and are on length scales from nanometers to micrometers. When considered across these length-scales, primary factors in the spatial organization are cell membrane domains and the actin cytoskeleton, both also highly dynamic. A key challenge is to understand the interplay across these multiple scales, link it to the physicochemical basis of the conformational behavior of single molecules, and ultimately relate it to cellular function. Overall, our premise is that at these scales, cell signaling should be thought of not primarily as a sequence of diffusion-controlled molecular collisions, but instead transient, allostery-driven cluster re-forming interactions.

**Introduction**

More and more data confirm: in the cell molecules that share a function cluster. This holds for membrane rafts; for receptors; for molecules anchored in the membrane, such as the nanoclusters of the Ras protein [1, 2]; for cytoskeleton proteins and the adaptors associated with them; for multienzyme complexes such as the MAPK and E3 ubiquitin ligases; for 'cellular factories' (discrete locations in the cell where co-functional enzymes are concentrated and anchored, such as RNA polymerases in transcription [3, 4, 5], and hGH gene regulation [6]), and more. Increasingly, data also confirm: the clusters are dynamic. For example, clusters in the cell membrane change dynamically their protein and lipid composition and locations; factories change the enzymes' copy number and cofactors' composition, adapting to the specific cellular location and cell state; multienzyme complexes such as the E3 ligases change the components of the machinery, dependent on the cell cycle state and the environment [7, 8]. Together, these indicate that cluster composition, location, structure and molecular interactions vary with time. Clusters are organized at multi-scales, from nanometers-scale of multimolecular clusters within the cell to micrometer-scale in intercellular signaling in the immunological synapses [9]. Collectively, the emerging picture is of environment- and signal-controlled clusters, with fluctuating patterns at different size scales, bridging direct, and lipid-, RNA-, or DNA-mediated indirect, protein-protein interactions. Pre-organization, where molecules are spatially and dynamically pre-positioned for productive association, spans the long-scale intercellular patterns. It takes place in membrane rafts, where dynamic protein interactions can be lipid-mediated, and in the actin cytoskeleton, which controls the dynamic spatial organization and mediates long-range interactions. It also takes place in the numerous clusters consisting of tens or hundreds of molecules, as in the case of the Foxp3, which forms complexes of 400-800 kDa or larger with 361 associated proteins identified by mass spectrometry, ~30% of which are transcription factors [10]. Here, our central thesis is that the coordination of the activities and responses of the cell to its environment emerge from this pre-organization across the cell, at different length scales, ready to be deployed. The multiple copies of the clusters, whose composition is modulated by the cluster spatial location in the cell, optimize the coordination. The dynamic states of the cluster composition and structure suggest how signaling varies in time. Together, these provide a framework of a spatial organization of signaling cascades, where signaling proceeds through intermolecular interactions between and within these clusters.

While numerous papers have addressed the intra- and intermolecular signaling (only a representative set of these are cited here [11-97]), very few touched on the longer-range, across- and inter-cluster communication [9]. Key questions are how signaling proceeds across the clusters, and how their spatial, mechanical and chemical properties relate to the conformational equilibrium and signaling efficiency. Below, we suggest that the pre-organization of proteins and other biomolecules (lipids, RNA, DNA) in clusters and the coordinated cellular response imply that signaling does not proceed over long scales by diffusion-controlled molecular collisions; instead, signaling proceeds through a population shift mechanism of the proteins across dynamically pre-organized clusters. Chance interaction of macromolecules during three-dimensional 'random walk' diffusion in open space depends on their concentration and ability to move rapidly over long distances [98]. However, the sub-nanomolar concentration of growth factors and the low number of membrane-bound ligands that stimulate cellular responses suggest that proximal signaling molecules interact at low concentrations, at least during early stages after cell stimulation. Further, interactions between freely diffusing small-molecule substrates and enzymes are less influenced by crowding owing to a large difference in the size of solute and 'crowders' [99], suggesting that the crowded cellular environment still does not imply diffusion-controlled recognition. On the other hand, the clusters are transient, and freely diffusing molecules may shift to form new clusters. Thus, while both the population shift mechanism and the diffusion-controlled chance collisions mechanisms can co-exist and are not mutually exclusive, signaling is likely to be more productive in pre-organized states. Coupled with other factors, primarily the concentrations of the proteins, cofactors and metabolites, and membrane composition, these help understand how despite cellular complexity, coordination and effective response is achieved.

**A view of the cell and its representation**

A simplified representation of the cell is helpful. Network diagrams provide the cellular pathways, their components and their links, from the extracellular domains of membrane-spanning receptors, through the cytoplasm to the nucleus. They often also depict the kinds of reactions that can take place, when the signal proceeds (or inhibited). A prime example of the usefulness of such diagrams is the popular KEGG resource [100]. At the same time, from the organizational and signaling standpoint, cellular diagrams may be misleading, obscuring cell coordination [101]. These diagrams overlooked the kinase repressor of Ras (KSR), a key scaffolding protein in the MAP pathways; they omitted the positive loop of the inhibitor of apoptosis protein (IAP) protein, an important drug target; and the overlapping interactions of protein partners which cannot take place simultaneously, in the case MyD88 (myeloid differentiation primary response 88 protein). The prevalent schematic cellular diagrams are typically represented as nodes and edges; they may span the cell, or focus on segments, such as those related to specific systems or organelles. They are often modularized to highlight functional units. Within modules the proteins can be expected to be in spatial vicinity; which may not be the case between modules. Yet, in reality, for the cell to function, the module composition needs to change dynamically, and proteins from one module would need to interact-directly or indirectly- with proteins in other modules. This raises a number of questions such as how do signals propagate among modules? Signaling requires physical interactions; and evolution is unlikely to have cellular communication programmed in a way that requires these proteins to randomly diffuse across large distances in the cytosol (or organelles) to convey a signal. While a random process can place, particularly during basal expression or cluster dissociation/re-association across long distances, it is not expected to be productive and robust if the modules are far away. Indeed, examples of communicating far-away modules are hard to find. Cells are commonly perceived as highly organized and structured, with membrane-enveloped organelles and cytoplasm, and sequestered functional units either attached to the membrane or partitioned and localized by cytoskeleton proteins. Such a high level of organization does not appear compatible with cell signaling being dependent on micrometer scale diffusion-controlled process. Signaling involves a complex set of ordered events. It may originate from the extracellular domain of a membrane-spanning receptor, or from a small molecule diffusing through the membrane. It has to activate, get amplified, lead to pathways uniting or branching - often through some combination of post-translational modification events [102], and get transferred all the way to the nucleus to activate or repress gene expression. Long distance diffusion would hamper cell action: while the volume excluded by the cytoskeleton increases the crowding and thus the intermolecular association constants, diffusion is a stochastic process, questioning whether the cell can afford to have long-range diffusion-dependent signaling. Questions can also be raised with respect to intra-modular signaling: while the diagrams depict them as single copy of single proteins connected by edges, are all proteins present at all times? And, do these modules contain a single, or even unique and constant number of copies per module? And finally, are they in direct contact or is the contact mediated by other molecules? Such questions underscore the gap between simplified diagrams and cell coordination.

Cellular processes need to be regulated; and regulation requires efficiency. Below, we partition the cell into macroscopic (organelles, modules) and microscopic (the functionally-related molecules within these) levels. We suggest that at both macroscopic and microscopic levels the cell is pre-organized, including the communication between the two levels. Pre-organization does not imply an immobile behavior; the distinct intermolecular interactions fluctuate, forming and dissociating with short residence timescales. These short-lived interactions have sufficiently long duration for the signals to go through, which allows coordination and priming successive enzymes in catalytic pathways [103, 104]. In the membrane, signals can proceed *via* membrane-anchored molecules (for example through myristoyl, farnesyl or palmitoyl groups), lipids (including cholesterol), and membrane-spanning receptors. In the cytosol, signals can transmit through the large assemblies, such as the nucleosomes in the nucleus; and as we argue here, also through the structured cytoskeleton, which is similarly dynamic. In all cases, scaffolding proteins [103, 105, 106] which are sometimes overlooked in cellular diagrams are likely to play major roles. Scaffolding proteins do not communicate the signal passively; they can control it [103]. Such a pre-organized, yet dynamic view of signaling in the cell emphasizes efficiency which proceeds not via chance collisions; but via a *population shift mechanism among pre-organized, albeit highly mobile molecules, that is, allostery*. The landscapes of the clusters can be highly heterogeneous in size, composition and shapes, and this heterogeneity is also governed by the cellular environment. The signaling state is location-dependent and likely to relate to this heterogeneity. Membranes and cytoskeletal structures reduce the reaction space by one and two dimensions, respectively, and hence increase the probability that molecular interactions will occur. This is thought to be one reason for the large number of signaling molecules bound to membranes and filaments. However, on its own, the reduced dimensionality of membranes and filaments still only marginally increases the likelihood of interactions between individual molecules. Exploiting the multiscale spatial cell organization and the conformational behavior of biomacromolecules, organizes signaling and coordinates cellular response.

Below, to introduce concept we start with the relatively small system of the Ras nanoclusters; then the larger intercellular clusters of ephrin, integrin, and immunological synapses, and finally with the pan-cellular structures of the cytoskeleton. We view the vital role of membrane rafts in signal transduction in this framework, via allostery, and end with briefly touching on experimental methods that can used to test hypotheses that flow from the conceptual framework outlined in this review.

**Dynamic clusters**

Cluster types and sizes vary [107, 108]. Cluster size estimations depend on the definition: in Ras, sizes are based on copies of only the Ras protein; in Foxp3, associated cofactors are also included. On average, there are seven Ras molecules in a nanocluster regardless of the activation state [109, 110]; ten or more receptors compose a microcluster on the surface of T and B cells [111-113] and hundreds of molecules in the large Foxp3 clusters [9]. Clusters are short-lived and highly dynamic. Ras is anchored in the plasma membrane through its C-terminal lipophilic post-translational modifications (PTMs) and positive charges. There are three Ras isoforms, H-Ras, K-Ras and N-Ras, with different PTM lipid anchors, which result in different preferences for raft-like liquid-ordered and nonraft liquid-disordered membrane domains. The farnesylated and double-palmitoylated lipid anchor of H-Ras is predominantly in cholesterol-enriched ordered domains; the farnesylated and single-palmitoylated lipid anchor of N-Ras mostly localizes at the interface between the ordered and disordered domains and the farnesylated and polycationic lipid anchor of K-Ras prefers disordered domains [1, 113, 114]. Ras is a key protein in the MAPK/ERK pathway which communicates a signal from receptor tyrosine kinases (RTKs) on the cell surface to the nucleus. An extra-cellular mitogen binds to the RTK. Via a series of events this leads to Ras (a GTPase) activation by swapping GDP for GTP. Activated Ras binds and activates the Raf kinase. Raf phosphorylates and activates MEK. MEK phosphorylates and activates MAPK (a mitogen activated protein kinase). The dynamics of Ras nanocluster assembly and disassembly control MAPK signaling. It is well-established that Ras nanocluster formation is essential for the activation of the MAP kinase cascade by RTKs [115]. Clustered, but not individually distributed, Ras proteins recruit and activate their downstream effector, Raf [108, 116]. Recently, it was observed that BRaf inhibition enhances nanoclustering of K- and N-Ras, but has no effect on H-Ras. This is important for two reasons. First, it provides insight into why clustering is essential for Ras action; and second, it underscores the difference between N- and K-Ras versus H-Ras cluster organization and activation. Raf inhibitors drive formation of stable hetero (BRaf-CRaf) and homo (CRaf-CRaf) dimers. Thus, two Ras-binding domains in a homo or hetero Raf dimer are required for increased K- and N-Ras nanoclustering, which suggests that Raf dimers promote nanocluster formation by serving as crosslinks for Ras.GTP (the complex of Ras with GTP, the active state of Ras) proteins [117]. Crosslinks increase the fraction of K-Ras and N-Ras in their respective nanoclusters, and enhance the cooperativity between the joined Ras monomers. Both effects can result in an increase in MAPK signaling. Further, there is cross-talk between the mitogenic Ras/MAPK and the survival PI3K/Akt pathways. The increased MAPK signaling in BRaf-inhibited cells decreases Akt activation. This might also be at least partially understood in terms of the nanoclusters: the MAPK/Akt pathway crosstalk reflects a competition between the stabilized Raf dimers and p110α (the catalytic subunit of PI3K) for recruitment to Ras nanoclusters. As we reason below, the fact that the H-Ras nanoclustering is not enhanced by BRaf can be explained by its predominant location in cholesterol-enriched ordered domains.

From our conceptual standpoint, Ras clusters can also be viewed as containing their downstream associated molecules, such as Raf (and other Ras partners, like PI3K). This is in line with our premise that cell signaling should be thought of not primarily as a sequence of diffusion-controlled molecular collisions, but instead as cluster forming interactions. Clusters of transcription factor Foxp3 also contain downstream partners, including the GATA-3 which facilitates Foxp3 expression [10], as do signaling clusters of other pathways as observed by co-localization experiments [108, 116, 118, 119; 120]. The enhanced nanoclustering and Ras activation by stabilization of Raf dimers also argue for a role for allosteric coupling between Ras molecules. The crosslinked Ras molecules essentially function as dimers even though they are monomeric GTPases. This may cooperatively enhance allosteric activation of Raf by Ras. For GPCRs, allosterism across homo- or heteromers, whether dimers or higher-order oligomers, represents such an additional topographical landscape [121]. As a sideline, such effects may offer the opportunity for novel therapeutic approaches. Crosslinking of lipid raft domains by multivalent ligands or antibodies are known to stabilize transient nanodomains and activate associated signaling complexes [122, 123], further arguing that this could be a mechanism in protein nanoclusters, either between monomers, as in the case of Ras; or between higher order complexes. In the case of the kinase Lck, which phosphorylates the T-cell antigen receptor (TCR), super-resolution fluorescence microscopy based on single molecule detection quantification of the cluster sizes has clearly illustrated that Lck conformational states regulate their own clustering, with the open conformation inducing clustering and the closed conformation preventing clustering [124].

Lck is anchored to the plasma membrane by myristoylated and palmyloylated N-terminal residues. Ras in anchored via combinations of farnesyl and palmitoyl. Palmitoylation [125], and likely other lipophylic PTMs such as farnesyl and myristoyl, can act as spatially organizing systems, efficiently counteracting entropy-driven redistribution of palmitoylated peripheral membrane proteins, acting to segregate the attached proteins into membrane domains. Regulation of palmitate turnover rates by extrinsic cues can arise from changes in the accessibility of thiol or thioester groups on the substrate protein upon conformational changes of the protein. Membrane anchoring of the lipophilic post-translational modifications also perturbs the protein conformation, leading to a functional conformational change, as in the case of the Ras protein. Importantly, at the same time, the lipid environment is also perturbed, and the perturbation can affect neighboring protein molecules as well. A GPCR crystal structure identified three water clusters in the receptor, totaling 57 ordered water molecules, two cholesterols which stabilize the conformation and 23 ordered lipids one of which intercalates inside the ligand binding pockets [126]. Allostery in the GPCRs has been shown to be partly controlled by ions (like sodium), membrane components (such as lipids and cholesterol), and also water molecules. A possible role for cholesterol mediating signaling can also be seen from BRaf inhibition: as we noted above, while BRaf inhibition enhances K-Ras and N-Ras clustering, there is no effect on the H-Ras, which is predominantly in cholesterol-enriched ordered domains. In ordered membrane domains, cholesterol-mediated Ras signaling can already take place, abrogating the effect of Raf crosslinks. Overall, Ras nanoclusters illustrate our view of cell signaling as allosteric cluster re-forming interactions rather than diffusion-controlled molecular collisions, highlighting the role of population shifts in the conformational ensembles, either directly or lipid-mediated.

The spatial organization of transmembrane receptors in lateral homotypic, heterotypic *cis*-interactions and intercellular trans-interactions, is important for receptor clustering and association with signaling proteins. Lipid microdomains can modify the activity of transmembrane receptors by (positively or negatively) influencing the clusters. Clustering of transmembrane receptors and lipid-protein interactions are important for the spatial organization of signaling at the membrane [127]. Among the examples for homotypic receptor clustering in *cis* are the RTKs, the intercellular Eph receptors (Ephs) and their ligands, the ephrins. Cell proliferation, differentiation, migration and adhesion are critical processes in development. Orchestration of the signal transduction is by two membrane-anchored hub protein families: the Eph receptor tyrosine kinases and their ephrin ligands [128]. The pre-clustered ephrins form homooligomers; upon cell-cell contact these bind the Ephs with a 1:1 stoichiometry. Clustering continues by formation of tetramers, inducing conformational changes on both the receptor and the ligand. The Eph tyrosine kinase domains trans-phosphorylate each other which initiates signaling; however, recruitment of Src-family kinases (SFK) to ephrinA/B ligands and phosphorylation of the ephrinBs leads to reverse signaling. The tetramers can be further clustered in higher-order assemblies which regulate the mode and strength of signaling. Different modes of clustering, such as through the extracellular [129, 130] or the cytoplasmic [131, 132] domains can also take place. Concentration is a key factor: at low receptor concentration, pre-clustered ephrin ligands are required for Eph clustering. Above the threshold, free EphAs can cluster independently of their ephrin ligand binding. High-affinity Eph/ephrin assemblies that form at the sites of cell-cell contact and are required for Eph signaling initiation [133]. In ephrin type A receptor 2 receptor Tyr kinase the clustering and the micrometer-scale spatial translocation of the clusters can result in mechanical-force sensing [134].

Another example is provided by the integrin, a family of α/β heterodimeric cell-surface receptors and the major mediator of cell attachment to the extracellular matrix whose members are able to signal across the membrane in both directions. Integrins are often found in highly organized clusters on the cell surface [135]. Here too direct protein-protein interactions are not always needed and signaling can proceed via lipid molecules, as in the case of cholesterol mediating the G-protein coupled receptor (GPCR) activation above [126] and H-Ras. Long scale spatial organization patterns may also be the result of the accumulation of multiple smaller clusters of tens to hundreds of molecules, as in the case of the T-cell receptors (TCRs) [9]. Such clusters have been observed in the immunological synapse also for other molecules, including LAT [136], ζ-chain associated protein kinase (ZAP70) [137], SH2 domain containing leukocyte protein of 76kDa [138], CD28 [139] and CD2 [140].

**A structured, though dynamic, eukaryotic cell**

Eukaryotic cells are highly organized, with a network of membrane-enveloped organelles. An added level of structural organization is provided by the cytoskeleton which maintains the cell's shape; anchors organelles in place; helps in the uptake of external materials, and in the separation of daughter cells after cell division. It also helps in moving parts of the cell in processes of growth and mobility. The eukaryotic cytoskeleton is composed of microfilaments, intermediate filaments and microtubles. Microfilaments are linear polymers of actin subunits that generate force by elongation at one end of the filament and shrinkage at the other. They act as tracks for the movement of myosin that attaches to the microfilament. Intermediate filaments are more stable and heterogeneous. Like actin filaments, they maintain cell-shape by bearing tension. Intermediate filaments organize the internal structure of the cell, anchoring the organelles and serving as structural components of the lamina in the nucleus. Microtubles comprise of α and β-tubulin polymers. A large number of proteins are associated with these, to control the cell structure. The cytoskeleton provides the cellular skeleton in the cytoplasm. It was proposed to increase the level of macromolecular crowding by excluding macromolecules in the cytosol [141]. Cytoskeletal proteins interact with cellular membranes [142] and together with the organelles' membranes, and the endoplasmic reticulum, and additional scaffolding proteins, help in further organization of the cytoplasm and the organelles, to segregate and co-localize functional units, which can be seen as autonomous functional units that turn on-, off-, and transmit signaling cues.

The cytoskeleton is dynamic [143], and the Rho family GTPases are its master regulators. Beyond regulation of actin filament organization by Rho GTPases, RhoD has a role in the organization of actin dynamics. RhoD binds the actin nucleation-promoting factor WHAMM, which binds the Arp2/3 complex, and the related Filamin-A-binding protein FILIP1. WHAMM acts downstream of RhoD, and regulates the cytoskeletal dynamics. The major effects on cytoskeletal dynamics indicate that RhoD and its effectors control vital cytoskeleton-driven cellular processes. RhoD coordinates Arp2/3-dependent and FLNa-dependent mechanisms to control the actin filament system, cell adhesion and cell migration [144]. Cell surface dynamics depend on the orchestration of the cytoskeleton and the plasma membrane by Rho GTPases [145]. Nucleotide exchange factors and GTPase-activating proteins regulate the activity of Rho GTPases. In turn, the cell cycle machinery regulates expression of proteins in the Rho signaling pathways through transcriptional activation, ubiquitylation and proteasomal degradation and modulates their activity through phosphorylation by mitotic kinases [146]. This regulated dynamic landscape points to changes in local cytosol composition and excluded volumes over time and space over different length-scales. It also argues for fluctuations in the level of local macromolecular crowding, and underscores our tenet of the potential of the dynamic cytoskeleton in mediating signaling across the cell via allosteric interactions.

The cytoskeleton, including the microtubules and the actin networks, do not merely provide structural support, withstand mechanical stress, drive cell motility, and form tracks; they are active regulators of cell cycle, development and fate [147-149]. Cell-fate determinant and checkpoint proteins have high affinities for microtubules and microtubule-dependent organization of non-membranous components directs cellular function [150]. Due to the density of the microtubule network, the sequestered molecules can form distinct module-like environments [149]. The extent of the sequestration depends on the binding affinities and the microtubule network density. Motor protein-mediated binding, which leads to convective fluxes, further helps in arranging the spatial localization of the molecules against the concentration gradient. The effectiveness of the microtubule-mediated sequestration and spatial organization can be observed in the Drosophila syncytial embryo. There, germ plasm proteins translocate from the posterior embryo cortex onto the mitotic spindles, becoming concentrated at the spindle poles, via an almost leak-free transport process across 5~10 mm [149, 151, 152]. During this long journey toward the pole, they almost certainly fall off the microtubule; but they are expected to rebind given the tight confinement, or else, given the concentration gradient, they would diffuse in the cytoplasm. Thus, the microtubule network density should be sufficiently high to cause the biased concentration of dyneins and their cargoes on one pole [149]. This example can provide a possible mechanism for the asymmetric distribution of cell fate determinants during the asymmetric cell division and formation of distinct module-like environments. Under modest microtubule densities, only partial sequestration and delayed diffusion are observed [153]. Thus, dynamic alteration of the microtubule network architecture can achieve different spatial sequestration patterns and spatial localization of cellular components. The organelle-like microtubule network compartmentalizes the cytoplasm, limits random diffusion, facilitates directed transportation, and thus can produce differential spatial distributions of cellular components [149]. These also result in changes in the network morphology and density. These coordinated actions can take place since tubulin, actin, and dynein are all highly dynamic allosteric proteins [154-158].

Microtubules are also highly dynamic structures, and since they contribute to most cellular functions, they need to be regulated in response to extracellular and intracellular signals; however, the linkage between the diverse signaling pathways and the regulation of microtubule dynamics is still unclear. Modifications of the tubulin dimer, tubulin modifying enzymes, and microtubule-associated proteins are all directly involved in the regulation of microtubule behavior and functions [159]. Microtubules undergo a broad range of post-translational modifications including polyglutamylation, polyglycylation, carboxyterminal cleavage and acetylation, whose functions are still not entirely clear. Among these, the constitutive and the inducible Hsp90 isoforms bind to microtubules in a way that depends on the level of tubulin acetylation. Tubulin acetylation also stimulates the binding and the signaling function of at least two of its client proteins, the kinase Akt/PKB and the transcription factor p53 [160]. p21-activated kinase 1 phosphorylates tubulin cofactor B (which facilitates the dimerization of α- and β-tubulin) and plays an essential role in microtubule regrowth [161].

**Signaling in the cell membrane**

Lipid organization in the cell membrane plays a vital role in signal transduction. Lipid rafts are membrane domains, more ordered than the bulk membrane and enriched in cholesterol and sphingolipids. We suggest that membrane rafts are also dynamically pre-organized and mediate allosteric signaling. While diffusion in membranes is 2D (rather than 3D as in the cytoplasm), this reduction in dimensionality does not provide efficient or robust signaling. In the membrane too, cell signaling should be thought of not primarily as a sequence of diffusion-controlled molecular collisions, but as sequences of cluster re-forming allosteric interactions, which is optimized in segregated ordered rafts. When disordered, it is helped by crosslinks, as we have seen in the cases of N- and K-Ras. Cholesterol can both drive the formation of ordered domains within the plasma membrane of cells, and we have discussed above for H-Ras and GPCR, directly mediate cell signaling via allosteric propagation [162]. Lipid rafts can be viewed as signaling platforms with variable composition and organization tailored for specific pathways [163] that initiate at the cellular surface. They have been implicated in numerous signaling pathways [164], whose regulation is adapted to these rafts. Composition and organization are inter-related: membrane domains lacking cholesterol differ in their organization from the ordered, cholesterol-containing domains. The example of the Ras nanoclusters, with the different regulation patterns of N- and K-Ras as compared to H-Ras illustrates this point: the H-Ras is predominantly in cholesterol-enriched ordered domains; N-Ras mostly localizes at the interface between the ordered and disordered domains and the K-Ras anchors in disordered domains. These preferred locations reflect the patterns of their lipophylic post-translational modification anchors. Lipid rafts modulate signaling molecules involved in multiple pathways. These include the pleiotropic src kinases [165] which activate the PI3K-Akt signaling pathway; the epidermal growth factor receptor (EGFR), which associates with caveolins and is involved in diverse processes including cell cycle regulation, endocytosis, and the MAPK cascade [166]. They also relate to the Ras pathway discussed above. It is this segregated domain organization of signaling molecules that led to the concept of the signalosome. A signalome contains interacting components of signaling pathways (such as EGFR) embedded in lipid rafts. It is choreographed by scaffolding proteins, such as caveolins, through compartmentalizing and concentrating signaling molecules. Further, in the cytoplasm scaffolding proteins allosterically control the regulation of multienzyme complexes [103]; a similar role can be assumed by scaffolding integral membrane proteins, with further involvement of lipid (sphingolipids and cholesterol) molecules. A good example is the CD40 signalosome associated with cell growth in B cell lymphomas. The CD40 signalosome is anchored in lipid rafts. Dysregulation leads to constitutive activation of the NF- kappaB pathway [167]. Similar signaling organizations operate in neuronal systems, such as that involving estrogen receptor (ER), which relates to neurogenesis, neuronal differentiation, synaptic plasticity, and neuro-protection.

The association between the ER, insulin growth factor receptor (IGF-1R), Cav-1, and a voltage gated anion channel, VDAC are also in lipid rafts. The formation of this signaling complex is lipid raft-dependent. In Azheimer's Disease proteins identified by mass spectrometry were clustered into specific signaling pathways, which allowed an appraisal of which lipid raft signaling pathways may be altered, rather than changes in individual proteins. This systems biology approach indicated that, in lipid rafts, wild-type mice had higher activation of pro-survival pathways such as PTEN and Wnt/β-catenin, whereas 3xTg mice showed activation of p53 and JNK signaling pathways.

The spatial organization in the plasma membrane (2D compartment) and the cytoplasm (3D compartment) differ. In the membrane, there is a horizontal, translational diffusion rather than diffusion in 3D space in the cytosol, or in 1D in the cytoskeleton. The reduced dimensionality of membranes (and filaments) still only marginally increases the likelihood of productive interactions between individual molecules. Exploiting the membrane rafts organization and the conformational behavior of proteins and lipids, can pre-organize signaling. Pre-organization pre-positions the lipid components spatially at preferred sites with respect to the proteins, as seen directly from the X-ray crystal structure of the GPCR. As we noted above, the crystal structure indicated 23 ordered lipids one of which intercalates inside the ligand binding pockets and two cholesterols which stabilize the conformation, and the presence of ordered water molecules as well [126]. However, even in the 2D cholesterol-enriched membrane compartments, multiple copies of the proteins can interact as in GPCRs homo- or heteromers, dimers or higher-order oligomers [121] as well as mechanisms involving other signaling proteins in the cluster, such as Raf crosslinking Ras.

**Some experimental methods that to test the hypotheses that flow from our conceptual framework**

Below, we provide a few possible experimental methods that have been used in previous studies and that will likely be used in future studies that can test the hypotheses that flow from the conceptual framework suggested in this review. We highlight especially relatively new techniques such as the optogenetic protein clustering [168]. This method obtains rapid and reversible protein oligomerization in response to blue light, and is based on *Arabidopsis thaliana* cryptochrome 2. Cryptochrome 2-mediated protein clustering can modulate signaling pathway in a dynamic manner; and as such can offer a way to quantitatively investigate signal transduction dynamics. In particular relevant to our proposition, is its ability to study the role of oligomerization as a mechanism in cellular signaling. Moreover, here the oligomerization is driven by light. Optogenetic systems are capable of accurate, dynamic control of signaling pathways through light-mediated protein heterodimerization and homodimerization. Light is an allosteric effector. This method provides a promising protein clustering system to target the fundamental higher level of signaling in the cell. Beyond proteins and small oligomers, it allows studies of signaling within and across clusters. Here the authors demonstrate its power by photoactivating the β-catenin pathway, and the RhoA GTPase.

Additional methods include super-resolution microscopy that enabled the characterization of TCR-dependent signaling clusters [169]. This method permitted the study of signaling microclusters at the single molecule level with resolution down to approximately 20 nm. It has further helped to characterize the size distributions of signaling clusters at the plasma membrane of intact cells. This method discovered dynamic and functional nanostructures within the signaling clusters, as predicted by our conceptual premise. Photoactivated localization microscopy (PALM) and direct stochastic optical reconstruction microscopy (dSTORM) followed the Lck ensemble distributions on the molecular level, and observed that they were controlled by the Lck open/closed conformational states [124]. The super-resolution fluorescence microscopy based on single molecule detection enabled quantification of the cluster sizes. Recent innovations in live-cell imaging at the sub-micrometer scale and object (particle) tracking also help observe signaling complexes and clusters and examine their dynamic character [98]. These allow addressing in greater detail the higher-order organization of signaling molecules in living cells. Fluorescence microscopy techniques are being suited to studying sub-micrometer signaling assemblies.

In addition, proteins identified by mass spectrometry were clustered into specific signaling pathways, which allowed evaluation of lipid raft signaling pathways and how these can be altered Alzheimer's Disease, rather than changes in individual proteins [170].

**Conclusions: dynamic interactions and conformational biasing across the cell**

Here we described our view of the spatial structure of cell signaling systems. Signals propagate through interactions; chief among these are between proteins. The hallmarks of protein-protein interactions resemble those present within protein cores [171, 172]. The protein interaction network spans the cell [173, 174]; some associations are long-lived, others take place over short time scales. The network is organized, and the interactions cooperative [175]. The allosteric signals propagate through these, traversing single molecules, their associations, clusters, and the cell. Clusters, with hundreds of molecules and varied compositions, are pre-organized and dynamic with tight binding occurring as the allosteric signal goes through. Their formation may be helped by other molecules, such as lipids and cholesterol [176], RNA or DNA. Ligand binding, or post-translational modifications, lead to conformational perturbation. To minimize the local frustration, conformational reorganization takes place. The consequent signal propagates, elicits conformational and (or) dynamic changes in far-away binding sites, leading to the specific selected recognition. The changes are ligand specific: different ligands will cause different conformational changes. Such scenario [103, 104, 177] can provide insight into coordinated cellular response.

While here we differentiated among the cellular organizations according to their scales, in reality there is a continuum of protein spatial organizations, from molecular complexes to domains and clusters, to the cytoskeleton; from cell-to-cell interface, to the membrane to the cytoplasm and the organelles. Such multi-scale organization across different levels can feed back to regulate specific proteins, and collectively cell signaling. The fundamental premise of Systems Biology is that system dynamics gives rise to cellular function [178]. All processes during cell life, including growth, differentiation, division, and apoptosis, are temporal; and they can be understood only in terms of dynamics; dynamics within- and among- modules, provide the clue to coordinated functional control. And within this framework, coordination is governed by a conformational biasing mechanism, that is, population shift. Population shift is the origin of allostery; it is the means through which action at the surface of one protein can be expressed by another, far away [96]. As the signal proceed, through short- and long-lived molecular interactions, mediated by proteins, lipids, RNA and DNA, it may get amplified or quenched, depending on other allosteric events along its long journey. And within this framework, efficient coordination exploits dynamic, linked, pre-organized clusters, spanning the cell.

To conclude, in 1998 Bray envisioned the optimal signaling module as being composed of membrane-bound upstream complex (in this case a cluster) and freely diffusing downstream regulatory molecules [179]. Here we suggested a modified view: rather than freely diffusing molecules, transient pre-organized and inter-connected clusters which span the cell, with signaling taking place via dynamic conformational population shifts. We reason that this may well be the efficient, robust and controlled signaling system embraced by evolution.

**Acknowledgements**

I thank the NCI group, Hyunbum Jang, Buyong Ma, Chung-Jung Tsai for many discussions over the years. This project has been funded in whole or in part with Federal funds from the National Cancer Institute, National Institutes of Health, under contract number HHSN261200800001E. The content of this publication does not necessarily reflect the views or policies of the Department of Health and Human Services, nor does mention of trade names, commercial products, or organizations imply endorsement by the U.S. Government. This research was supported (in part) by the Intramural Research Program of the NIH, National Cancer Institute, Center for Cancer Research.